# Recent Trends of Measurement and Development of Vibration Sensors

**Himanshu Chaurasiya**

**Assistant Professor, Electronics & Communication Engineering Department,**
A.S.E.T., Amity University, Noida, U.P. 201303 INDIA
**hchaurasiya@amity.edu**

**Abstract**
In recent trends, Sensors are devices which monitor a parameter of a system, hopefully without disturbing that parameter. Vibration measurement has become an important method in mechanical structural product's research, design, produce, apply and maintenance. Vibration sensor is more and more important as key devices. Nowadays, with the development of computer technology, electronic technology and manufacturing process, a variety of vibration sensors have come forth in succession.
*Keywords: Development trend, Present situation order, Sensor, Vibration measurement.*

## 1. Introduction

Vibration is one of the most popular phenomena that exists in our daily life, which is everywhere and at all the time. Vibration is generated as a result of mechanical disturbance from sources such as music/sound, noise, engine, wind and many more. Detection of vibration is an important sensor technology for monitoring the operation of machines, bridges and buildings, warrant of security, prediction of natural disasters and more. As we know, the vibration sensor testing technology has been developed gradually from early last century. With scientists' exploring and researching, and accordingly the test methods and the types of sensors are evolving and maturing. Vibration measurements usually include vibration displacement, velocity, acceleration and others' measurement, usually, the device that converse the vibration into the electrical is called as vibration sensor [1]. Especially in recent years, vibration measurement has become an important method in mechanical structural product's research, design, produce, apply and maintenance [2]. Thus, a variety of vibration sensors made by the effect of physical have drawn more and more attention, with the development of computer technology, electronic technology and manufacturing process, a variety of vibration sensors have come forth in succession in order to using in different areas.

## 2. Present Situation of Vibration Sensors

According to the principle of vibration sensors, several current vibration sensors are described which are used widely in basic principles and features; in the end the development trend of vibration sensors was viewed. Vibration sensor has many types; its basic measurement principle is shown in Figure 1. Vibration sensor detects the vibration parameter of objects through its mechanical structure, and converting the vibration parameter into the electrical signal by physical effect to achieve transferring the non-electrical signal to electrical signal. Vibration sensor separates into displacement (amplitude) sensor, speed sensor and acceleration sensor according to the measured vibration parameters. Because of the displacement, the velocity and the acceleration can be translated into each other in the way of simple calculus; the three kind sensors can universal sometimes. Currently, according to different methods of detecting vibration, vibration sensors with different kinds of physical effects are invented, which are used widely in the following categories.

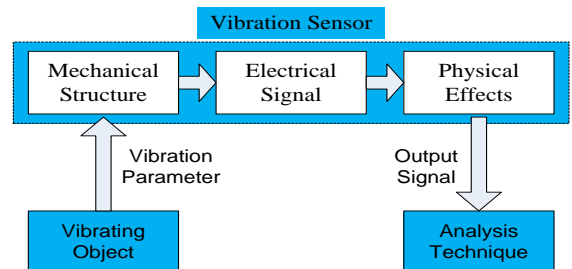

Fig. 1 The basic measurement principle of vibration sensor

### 2.1 Inductive Sensors

An inductive sensor is an electronic proximity sensor, which detects metallic objects without touching them. Inductive sensors are based on electromagnetic induction, use self-inductance coil or mutual inductance coil to achieve detecting the electrical signal which is converted from the vibration [3]. Inductive sensor principle structure is shown in Figure 2.

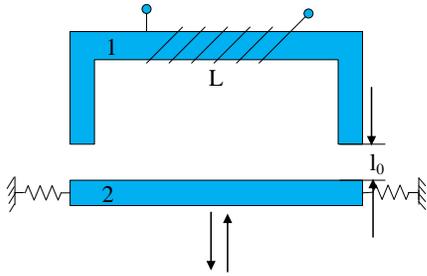

Fig. 2 The principle structure inductive sensor

Where, 1 is the fixed Iron, 2 is movable armature. N circles of wire are circled on the fixed Iron. According to the knowledge of the magnetic circuit, when the fixed Iron works in the non- saturation, then

$$L = \frac{N^2 \mu_0 A_0}{2l_0} \quad (1)$$

Where, $\mu_0$, $A_0$, $l_0$ respectively are the magnetic permeability, equivalent cross-sectional area and length of the gap, When the vibration of an object causes one changes of the area or thickness or the magnet length between the fixed Iron and the armature, the inductance changes, and measuring the change of inductance in order to achieving measuring vibration.

The prominent features of inductive sensor conclude structure simple, reliable, which have high accuracy, zero stability, and great output power, and so on. Our inductive sensors are capable of detecting any ferrous metal. The disadvantages conclude the sensitivity, linearity and range are restricted by each other, which cause to be not suitable for measurement of high-frequency dynamic signal [4].

2.2 Piezoelectric Sensors

Piezoelectric sensor is a typical sensor, which can generate electricity itself. It bases on the piezoelectric effect of some piezoelectric materials. When it suffers vibration, the surface of piezoelectric materials will produce electric charge. After amplification by the voltage amplifier or the charge amplifier and impedance conversion this electric charge becomes the power output in direct proportion to the outside force the sensor suffer. The goal of measuring vibration parameters which is non-electricity signal can be realized. Piezoelectric sensor is mainly used for measuring the dynamic force and acceleration. Piezoelectric sensor principle structure is shown in Figure 3.

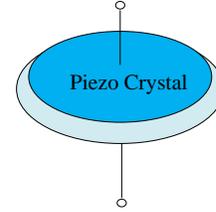

Fig. 3 The principle structure piezoelectric sensor

The prominent features of piezoelectric sensors conclude no-moving-parts, and the width of frequency band, high sensitivity, the high ratio of signal to noise, and the structure is simple, reliable, light weight. The disadvantages conclude resonant frequency, vulnerable to interference from the external environment, high output impedance, weak output signal which requires amplification through the amplifier circuit and detection by detecting circuit. Currently, with the rapid development of electronic technology, and the accompanying secondary instrument and cable of low noise, high insulation resistance and small capacitance enable the application of piezoelectric vibration sensors used more widely[5].

2.3 Magnetic Sensors

Magnetic sensor is also known as electric sensor, it transforms the vibration parameters into the induced electromotive force. It is a transforming sensor which changes mechanical energy into electricity energy. Magnetic sensor bases on the principle of electromagnetic induction, which is shown in Figure 4. According to the law of electromagnetic induction, the induced electromotive force in the coil is proportional to the magnetic flux rate of change, that is

$$e = -N\frac{d\Phi}{dt} \quad (2)$$

Where, N is numbers of turns for the coil, $\Phi$ is flux surrounded by the coil.

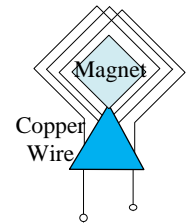

Fig. 4 The principle structure magnatic sensor

The magnetic flux rate of change relates to magnetic field strength, magnetic reluctance, the velocity of the coil. When the vibration causes one changes of these factors, it will change the induced electromotive force in the coil. By measuring the induced electromotive force changes can realize the purpose of measuring vibration.

The prominent features of Magnetic vibration sensor conclude large output signal, simple post-processing circuit, and powerful anti-interference ability. The disadvantages conclude relatively complex and large structure. At present circuit-correction methods can be used to reduce the test frequency of magnetic vibration sensor, as well as in low frequency vibration test [6].

### 2.4 Capacitive Sensor

Capacitive sensors detect anything that is conductive or has a dielectric different than that of air. Capacitive sensor is the instrument which changes the non-electricity-signal parameters into electrical capacity, then change the electrical capacity into voltage or current, using the principle of capacitor. In the vibration field, capacitive sensors generally divided into two types, the variable-clearance type and the variable-area type. Figure 5 is the structure diagram of variable-clearance sensor.

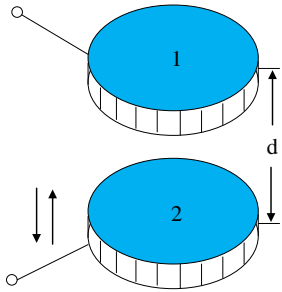

Fig. 5 The structure diagram of a variable-clearance sensor

Where, 1 is the fixed plate, 2 is the movable plate. When the vibration of an object cause the moving plate moves up for $\Delta d$, the capacitance increment is

$$\Delta C = \frac{\varepsilon A}{d - \Delta d} - \frac{\varepsilon A}{d} = C_0 \bullet \frac{\Delta d}{d - \Delta d} \qquad (3)$$

Where, $C_0$ is the initial capacitance value when the pole pitch is d.

As can be seen from the above equation, the change of capacitance relates to the displacement of the moving plate, and when $\Delta d \ll d$, we can approximate that $\Delta C$ is linear relationship with $\Delta d$. So by measuring the changes of capacitance and thus measure the vibration parameters. By measuring the induced electromotive force changes can realize the purpose of measuring vibration. More and more design engineers are selecting capacitive sensors for their versatility, reliability and robustness, unique human-device interface and cost reduction over mechanical switches. Figure 6 is the structure diagram of variable-area sensor.

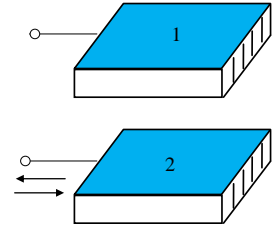

Fig. 6 The structure diagram of variable-area sensor

Where, the 1 is fixed plate, 2 is movable plate. When the vibration causes the changes of public area between plate 1 and plate 2 so that causes the changes of capacitance. When the capacitance between the areas covered A becomes A′, and then the variable capacitor is

$$\Delta C = \frac{\varepsilon A}{d} - \frac{\varepsilon A\prime}{d} = \frac{\varepsilon(A - A\prime)}{d} = \varepsilon \bullet \frac{\Delta A}{d} \qquad (4)$$

Seen from the above equation, $\Delta C$ is linear relationship with $\Delta A$, so by measuring the change in capacitance and thus measure the vibration parameters. Linear vibration displacement can be measured by the variable-clearance type capacitive sensor; Angular displacement of torsion vibration can be measured by the variable-area type capacitive sensor.

The prominent features of the capacitive sensor conclude high resolution, wide measurement range, high precision, short dynamic response time, suitable for online, dynamic measurements and non-contact measurements [7]. The disadvantages conclude small measuring-range, high output impedance, having parasitic capacitance, low-grade anti-jamming ability, and its measure is vulnerable to electrical medium and electromagnetic fields [8]. Now, with the in-depth research of capacitive sensor measurement principle and structure, and the development of new circuit, new materials, new processes, some of its shortages gradually are being overcame. The accuracy and stability of capacitive sensors are increasing and used more widely in non-contact measurement field.

### 2.5 Optic Fiber Sensor

Generally by the optical fiber sensor, laser and light detector composed of the three parts of Optic fiber sensors. According to the different Operating Principle of Optical fiber sensor can be divided into functional and non functional. The former is the use of the characteristics of the fiber itself, and use the optical fiber as the sensitive components. The latter is the use of other sensitive components to detect changes of the measured physical quantity; just optical fiber is used as transmission medium to transport the optical signal from distant or inaccessible location of. In practice, the optical fiber as the sensitive components of vibration information directly is difficult to separate the impact of changes from other physical quantities, therefore, the non-functional optical fiber vibration sensors is widely used in the field of vibration detection, where the basic principle is the use of other sensitive detect changes of the measured physical quantity, and the light parameter is modulated by sensitive components [9]. Figure 7 is a phase-modulated optical fiber vibration sensor schematic diagram.

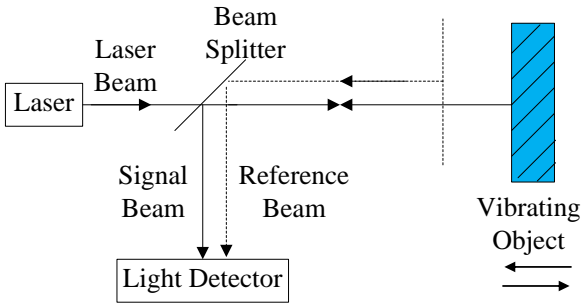

Fig. 7 The schematic diagram of a phase-modulated optical fiber vibration sensor

The vibrating object change the relative phase of signal beam and reference beam, which result in a phase modulation, and by demodulation and detection the phase modulation, you can get the corresponding vibration amplitude.

The prominent features of the optical fiber sensor conclude their ability to be lightweight, very small size, high sensitivity, fast response, resistance to electromagnetic interference, corrosion resistance, electrical insulation, soft bend, suitable for long-distance transmission, and easy to connect with the computer and make telemetry network with fiber optic transmission systems, especially can long-distance vibration in harsh industrial environments. Practice has proved that it has high sensitivity and reliability of persistent work, which can detect the vibration amplitude from 10-12 meters and can be used to three-dimensional vibration measurements.

The disadvantages conclude the narrow range of measurement frequency, high cost and unfamiliarity to the end user. Therefore, the optical fiber vibration sensor has the broad value of further research and development.

2.6 Photoelectric Sensor

Photoelectric sensors first non-power vibration parameters change into the changes of light, and then through the use of photoelectric effect of the optoelectronic device, light signals changed into electrical signals and finally achieve the intention of converting the changes in the vibration parameters to the changes in power [10]. Figure 8 is a principle schematic diagram of the vibration sensor based on laser interferometer Principle.

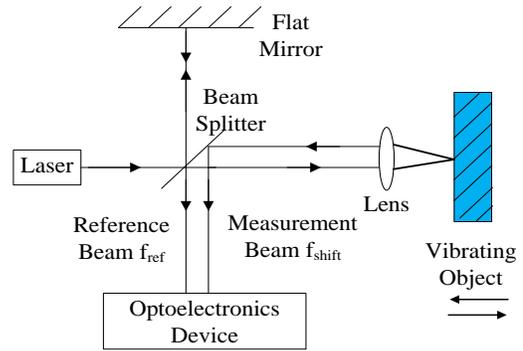

Fig. 8 The principle schematic diagram of the vibration sensor

Laser beam emitted by the laser divide into the reference beam and measurement beam with beam splitter. Reference beam reflected by the flat mirror, and then through the beam splitter again, finally direct into the optical components, the measurement beam is focused by the lens after shooting the measured vertical vibration of the surface, when the vibrating object vibrate in the direction of laser beam, due to the Doppler effect occurs frequency shift of the frequency $f_{shift}$ of the reflected light is collected by the objective lens deflection shooting at the beam splitter optoelectronic device, and with the frequency $f_{ref}$ of the reference beam to interfere in the photosensitive surface to achieve mixing. The optical components convert the optical frequency difference obtained by mixing to be electrical signal, and vibration amplitude and vibration frequency can be available through signal processing.

The prominent features of the photoelectric sensor conclude high resolution, high precision, fast response, noncontact, etc., especially the photoelectric sensor, which using laser as the light source, making use of laser interference, diffraction, and high precision of the developed high precision measurements in optical vibration sensor vibration field more and more attention. The disadvantages conclude the distance measurement is limited, and optoelectronic device characteristics impact by environmental interference, the subsequent processing circuit complex and difficult application. Currently, because accuracy of photoelectric sensor can reach the nanometer grade, but it is too hard to used in practical application of engineering, so the research of photoelectric sensor have became one of hottest topics in the modern international world.

## 3. Prospect of the Vibration Sensor

In engineering, the applications of vibration sensor are widely, so it caused by a high degree of importance about its research and development in the world. At present, with the development of science and technology, the shortcomings of vibration sensors continue to be overcome, measurement accuracy and increasing the sensitivity range of applications are increasingly being used, the developing prospect of vibration sensor mainly are the following aspects:

### 3.1 Performance of real-time measurement

Now, with industrial production and machinery manufacturing developing into the direction of high precision, people pay more attention hazards of high frequency and low-frequency micro-vibration, which requires vibration sensor to measure vibration band width, the resolution must be high, faster response time, higher sensitivity, meet the ask of real-time vibration measurement, control high-frequency micro-vibration and low-frequency micro-vibration hazard.

### 3.2 Integrated, intelligent and modular

Exploiting the experiences of integrated circuits miniaturization, reducing vibration sensor in weight by the sensor technology miniaturization of the hardware system, improving their processing speed and reducing the effect to the measured vibration parameters of object [11]. While taking advantage of multi-information fusion technology, combined the vibration sensor technology with microprocessor computing, the of both detection and judgments, and wireless remote control and information processing capabilities, making vibration measurements to achieve all digital intelligent, modular measurement.

### 3.3 Preferable Anti-jamming ability, low-impedance output

Vibration sensors on a variety of outside interference has great influence on measurement accuracy, and need to improving its anti-jamming capability, especially in the development of laser characteristics for the principle of noncontact vibration sensor to measure the optical path and installation of structural simplification, anti-interference ability high precision, convenient engineering applications for a wide range. At the same time, low-impedance output, reducing the requirements for subsequent processing circuitry.

### 3.4 Preferable environmental adaptability, remote Measurement

Improve the reliability of vibration sensor, so it can work in harsh environments and high-intensity vibration effectively and maintain the accuracy of measurement, remote measurement can be overcome barriers to engineering applications.

## 4. Conclusions

With the development of sensor technology, vibration sensors in the form will be varied, through the extensive application of electronic technology to achieve high overall performance further. Vibration sensors will play an increasingly important role in scientific research and automate production process, and its development will profoundly affect the development of national economy and national defense science and technology.

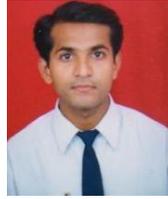

Himanshu Chaurasiya received the Bachelor of Technology degree in electronics & communication engineering from Uttar Pradesh Technical University, Lucknow, INDIA in 2006 and Master of Technology in Digital Communication from Bundelkhand Institute of Engineering & Technology Jhansi INDIA in 2009. He is currently assistant professor of electronics & communication engineering department, at A.S.E.T., Amity University Noida, U.P., INDIA. He has been actively involved in teaching & research. His research interest includes optical fibers communication, sensors and signal processing specially speech processing. He has life member of ISTE, IACSIT and IAENG. He had written a book titled Implementation of Automatic Speech Recognition using Soft Computing under the LAP Germany. His technical papers have been published at International and national level journals and conferences.